# TMI! How Knowledge Platforms Tame the Information Overload and Advance Global Development Through Technology

Rob Goodier
Engineering for Change
New York City, NY, USA
editor@engineeringforchange.org

Iana Aranda
Engineering for Change
New York City, NY, USA
iana@engineeringforchange.org

Laura MacDonald
Centre for Affordable Water and Sanitation Technology
Calgary, AB, Canada
lmacdonald@cawst.org

## ABSTRACT

Finding reliable data to inform decisions about technology for global development remains a challenge. Easily accessible "Knowledge platforms" are a way to curate and standardize information about technology for development. Three collaborators, Engineering for Change, the Global Alliance for Clean Cookstoves and the Center for Affordable Water and Sanitation Technology (CAWST) are working together to create platforms to serve the global development sector.

Such platforms could be the first step in making decisions about how to solve a problem that needs a technology-based solution. They could motivate manufacturers to demonstrate compliance with quality standards, and they could encourage independent research into each product's impact. Years of development experience worldwide have yielded a set of best practices to increase the life of a project. These platforms clarify those practices to improve quality and reduce waste.

As the platforms mature, mining them for data could identify trends that influence the entire technology ecosystem. Those could include decisions about performance targets, future innovation, pricing, compliance with regulations, and funding priorities.

## INTRODUCTION

In 2008, with his Brown University diploma fresh off the press, Drew Durbin founded SolarCycle and tried to sell solar cooking ovens in Mozambique. SolarCycle converted plastic bags into reflectors that focus sunlight on a cooking surface, a plan that merged waste recycling with solar power.

Solar ovens cook cleanly, the fuel is free and SolarCycle's plan seemed solid.

But then Durbin and his team began to put it into action. They found that they could not find a facility in Mozambique that would convert bags into reflectors, then they realized that solar ovens do not sell themselves. The idea would need the support of an educational marketing campaign. And even after the sales pitch, not everyone was convinced. Cooking with the sun depends on the weather and the time of day, and it also requires an outdoor kitchen. For families that wanted a hot dinner at home after dark, this type of solar oven was not impractical.

"The cooker is something that happens over and over again to a lot of people. They get really excited about this technology but it doesn't gain traction," Durbin says.[1]

Solar cooking has a nuanced history in the developing world and some efforts to distribute them have been successful.

But Drew Durbin's problem is a common one in global development. Ideas that shine among design teams and investors from developed countries fizzle when they meet the reality of factories, store shelves and homes in developing countries.

"Solutions too often reinvent the wheel rather than building on robust platforms, infrastructure, and shared services. Applications and services designed thousands of miles from their use environment fail to meet user needs. The creation of duplicative tools and systems has made data difficult to access and use for decision-making. This is an inefficient use of scarce resources. We must do better, both to fulfill our own mandates and, critically, to deliver to the best of our ability for the people we serve" says Ann Mei Chang, Executive Director, U.S. Global Development Lab USAID.[2]

In 2008 it might have been hard for Durbin to find research on solar ovens that could have prepared him for SolarCycle's hardship. Today that information is available, much of it in a piecemeal fashion scattered throughout the Internet. Until now. Engineering for Change, The Global Alliance for Clean Cookstoves and the Center for Affordable Water and Sanitation Technology (CAWST) join a handful of organizations working to collect information about global development technology into easily accessible "knowledge platforms." These platforms are repositories of standardized data on the technology that meets basic needs in areas where resources are constrained.

## 1. THE RATIONALE FOR NEW RESOURCES

Engineers and others in the field have often asked three questions about technology for low-resource regions. They are, "Which solution is appropriate for a particular context?" "What has scaled and what has stalled?" and "Does this solution meet the need?" And if those are answered, another pair of questions that are often asked are, how do I build a thing and where can I find technical support? Engineering for Change, CAWST and others have responded to those questions and others with their respective knowledge platforms.

## 1.1. ENGINEERING FOR CHANGE: SOLUTIONS LIBRARY

Engineering for Change created its Solutions Library to help answer all five commonly asked questions. The Solutions Library is a central information source that could increase efficiency and reduce wasted work in global development.[3]

Starting in 2012, Engineering for Change convened international teams of development engineers with backgrounds in the private sector, non-profit and academia. Those experts collaborated to define and prioritize data related to essential technology for underserved communities. To date, the Library has codified 403 products and services in eight categories, Energy, Agriculture, Health, Water, Sanitation, Information Systems, Transport and Housing.

The work is ongoing, carried out by cohorts of research fellows, guided by a global network of expert advisors and enabled by collaboration with organizations that are equally committed to data quality such as CAWST and the Alliance for Clean Cookstoves. The result is a living database of un-biased, codified product reports and comparisons of products for designers, manufacturers and implementers. It is a due-diligence effort that funnels the wide variety of units, measurements and research into a standard format.

These resources could be the first step in making decisions about how to solve a problem that needs a technology-based solution. When the problem is how to store grain after the harvest, for example, the Solutions Library allows for side-by-side comparisons of the Interlocking Stabilized Soil Brick Granary, the Zero-Emissions Fridge for Rural Africa made of woven bamboo and clay, and Purdue Improved Cowpea Storage Bags. When the problem is how to charge mobile phones, the Library can compare 28 devices including solar panels, solar home systems, kiosks, stoves that charge phones a solar suitcase and even a toilet that generates electricity from urine.

## 1.2. CAWST: AN ONLINE KNOWLEDGE PLATFORM FOR THE WATER, SANITATION AND HYGIENE SECTOR

CAWST's digital platforms and in-person support can answer questions about which solution is appropriate and will it meet the need. But the organization's strength is in its technical assistance and construction guides. CAWST began work in 2001 as an in-person educational, training and consultation service to those who work in water, sanitation and hygiene (WASH) in developing countries. Over the years as CAWST grew, along with demand from its clients for technical support. In 2012, CAWST turned to the Internet to meet the challenge. The organization created Web sites and webinars, including a WASH Education and Training resources website, a Biosand Filter Knowledge Base to offer technical details about the low-cost water filtration technology, a Household Water Treatment and Safe Storage Knowledge Base, a mobile application for offline users, live chat for online expert consultation, a WASH e-library, and a series of online training sessions.

CAWST's leadership has noticed the same need for better coordination among its constituency in water and sanitation that Engineering for Change, USAID, the Global Alliance for Clean Cookstoves and others in the development sector have expressed.

With advances in digital tools that include communication technologies, social media and online mapping, that coordination is now possible.[4]

## 1.3. THE GLOBAL ALLIANCE FOR CLEAN COOKSTOVES: CLEAN COOKING CATALOG

To help practitioners assess which stove will be appropriate in each context, the Global Alliance for Clean Cookstoves launched an online resource called the Clean Cooking Catalog in 2013.[5] The catalog is a database of 322 cookstoves, 14 fuels and performance data from 647 tests of clean-burning stoves used and made around the world. These stoves are created to replace open cooking fires and stoves that smoke excessively, burn fuel inefficiently or both. The catalog standardizes the information that is available about stoves by including data on specifications, emissions, fuel efficiency and safety, all based on laboratory and field tests.[6]

## 2. STANDARDIZED INFORMATION AS A CATALYST FOR PRODUCT IMPROVEMENT

The E4C Solution Library's descriptions of products include the results of testing (if it was carried out), and outbound links to research literature (if it exists). The absence of that data could be viewed as a warning to potential buyers. When products designed for the developing world demonstrate a lack of compliance with quality standards, whether the standards are international or regional, questions arise as to whether the standards or compliance frameworks exist.

Few products have been tested independently, either to ensure that they are not defective or to quantify the impact that they can have among their users. Needless to say, those omissions can lead to wasted investment at best, and they can be life-threatening at worst. Medical devices, water treatment, food storage, sanitation and other products can save lives if they work properly. With a consistent and defined data framework the missing research may be more glaring. The hope is that the manufacturers will take notice and work to fill gaps in testing their products.

## 2.1. EXAMPLE: CAWST'S BIOSAND FILTER KNOWLEDGE BASE

One of CAWST's online resources, the Biosand Filter Knowledge Base, organizes and presents information useful to researchers and practitioners interested in the filtration technology.[7] The filter itself was in use by 5 million people in 55 developing countries as of the Fall of 2015. And the knowledge base offers answers to frequently asked questions about the filter, summaries of research papers, independent evaluations and projects that the organization's clients are carrying out.

## 2.2. EXAMPLE: THE CLEAN COOKING CATALOG

The Global Alliance created their catalog to promote the adoption of international clean cookstove standards.[8] To track their progress and impact, the alliance set definitions for "clean" and "efficient" in line with tiered performance guidelines established in the ISO International Workshop Agreement in 2012. The tiered system ranks stoves by their fuel efficiency and indoor and overall emissions. The framework is dynamic, updated as new evidence is available and new progress on standards development is made.[9]

## 3. NEW ANSWERS THROUGH DATA MINING

Data mining will become possible as the Solutions Library, the Clean Stove Catalog and CAWST's knowledge bases expand. Trends that emerge could influence the entire technology ecosystem, from decisions that designers make about future upgrades, decisions that manufacturers make about compliance and distribution, and those that funders make about which solutions to invest in.

For example, CAWST knows that its Biosand Filter Knowledge Base has 1000 users in 68 countries. A survey in April 2016 found that 61 percent of the users are practitioners, 41 percent are in academia as researchers or students, and 46 percent said they had used the resource to help build a biosand filter.[4]

CAWST can already apply basic analysis to its data to quantify its own educational impact. The April 2016 survey found that 43 percent of the people who used its Biosand Filter Knowledge Base did so to train biosand filter technicians. That amounts to an average of 85 technicians per respondent.[4]

This kind of information represents the start of a potentially vast and helpful resource.

As a hypothetical example, cross-referencing geographical data with distribution to date and performance could reveal preferences in a certain region. This kind of information could be useful to hardware entrepreneurs who could determine which percentage of households in rural Kenya prefer high quality solar lanterns and are willing to pay more for them.

Those data could also reveal the distribution schemes that are likely to be successful in a given area. Unlike in a strictly capitalist market, the means of distribution vary widely in the field of global development technology. Products can sell on an open market, but they might also be financed, subsidized or even given away freely at the expense of a government or other donor. Discussion over which method is preferred can be informed with standardized information about each product.

## 4. CONSERVING RESOURCES THROUGH EDUCATION

Years of development experience worldwide have yielded a set of best practices to increase the life of a project and mistakes that can cost time and money. Knowledge platforms can elucidate those practices to improve the quality of technological interventions in impoverished regions and reduce waste.

Well-meaning people and organizations tend to make a small set of similar mistakes that doom their investments in technology. These common mistakes can include the development of a new product without first researching the needs of the community for which it is intended, installing infrastructure in a community without assuring a means of maintaining and repairing it, or introducing new hardware to a community and then leaving with no commitment to long-term follow up. Development professionals are familiar with those and other mistakes, and their field is rife with examples.

In a common scenario, a group in the United States sets out to help a community in a developing country with its water treatment. To do that, they identify a community, find or make a water treatment product and then install it. That amounts to a 'dead end' and it is often repeated, says Daniele Lantagne, a Civil and Environmental Engineering Professor at Tufts University in Medford, Massachusetts.

"This type of implementation does not take into account local water quality, water treatment technology selection, the community knowledge, attitudes, and practices around water, the desire (or lack of it) of the community for water treatment, and the necessary behavior change communication. It been shown again and again to be ineffective" Lantagne says.[10]

The knowledge platforms mentioned in this report have the potential to prevent these kinds of mistakes. Engineering for Change offers an introductory course in the design and delivery of global development technology, monthly webinars with experts, and expert interviews and in-depth investigation and analysis.

The ineffective water projects that Lantagne mentioned, for example, could benefit in part from a webinar on how to protect water infrastructure investments through long-term preventive maintenance.[11]

Likewise, a study by a team at Pennsylvania State University may give pause to early adopters who view 3D printing as a practical means of manufacturing prostheses in rural communities. They analyzed the impact of the method and suggested that the technology may need time to mature, and that the real problem may not be limb production, but the lack of trained prosthetists who can fit limbs.[12]

## 5. THE CHALLENGES THAT REMAIN

The work to standardize and harmonize information about technology designed for underserved communities has only just begun. Engineering for Change, the Global Alliance and CAWST have laid a foundation for conformity that we hope designers, manufacturers and funders will adopt. But product information remains, for the most part, heterogeneous. And it also remains siloed in disconnected Web pages, slide shows, books, reports and elsewhere. The result is limited data sharing that stymies learning and innovation on products that improve the quality of life for the people who most need it.

Building a supportive environment for data sharing requires continued investment in advocacy for rigorous knowledge development and analysis, combined with a commitment to real collaboration.

## 6. REFERENCES

1. Goodier, R. (2011, October 17). Local Production: The Case For and Against. Retrieved from Engineering For Change - http://www.engineeringforchange.org/local-production-the-case-for-and-against/

2. Waugaman, A. (2016). From Principle to Practice: Implementing the Principles for Digital Development. Retrieved from UNICEF Stories - http://www.unicefstories.org/wp-content/uploads/2013/08/From_Principle_to_Practice.pdf

3. Solutions Library. Retrieved from Engineering for Change - http://solutions.engineeringforchange.org/

4. MacDonald, L. (2016). An Online Space for Practitioners in the Water, Sanitation and Hygiene Sector. International Journal of Environmental, Chemical, Ecological, Geological and Geophysical Engineering, 10(5). Retrieved from http://internationalscienceindex.org/publications/10004426/an-online-space-for-practitioners-in-the-water-sanitation-and-hygiene-sector

5. Clean Cooking Catalog. Retrieved from the Global Alliance for Clean Cookstoves - http://catalog.cleancookstoves.org/

6. About the Catalog. Retrieved from Clean Cooking Catalog - http://catalog.cleancookstoves.org/pages/about

7. Biosand Filters Knowledge Base. Retrieved from CAWST - http://biosandfilters.info/

8. Standards. Retrieved from Global Alliance for Clean Cookstoves - http://cleancookstoves.org/technology-and-fuels/standards/index.html

9. How does the Alliance define "clean" and "efficient"? Retrieved from Global Alliance for Clean Cookstoves - http://cleancookstoves.org/technology-and-fuels/standards/defining-clean-and-efficient.html

10. Comments made for this report in an interview by email May 2016. See also: Goodier, R. (2012, November 14). Five Questions with Daniele Lantagne. Retrieved from Engineering for Change - https://www.engineeringforchange.org/five-questions-with-daniele-lantagne/

11. Future-Proofing Water Systems in Developing Countries: How to Protect Investment. Retrieved from Engineering for Change - https://www.engineeringforchange.org/webinar/future-proofing-water-systems-in-developing-countries-how-to-protect-investment-and-increase-success-through-preventive-maintenance/

12. Phillips, B. (2016, February 4). 3D-Printed Prostheses May Not Be Practical in Developing Countries, Yet. Retrieved from Engineering for Change - http://www.engineeringforchange.org/3d-printed-prostheses-may-not-be-practical-in-developing-countries-yet/